\documentclass[useAMS]{mn2e}
\usepackage{graphicx}
\usepackage{epstopdf}
\usepackage{amsmath}
\title[Exact analytical solutions for ADAFs]{Exact analytical solutions for ADAFs}
\author [Habibi, Abbassi, Shadmehri]{ Asiyeh Habibi$^{1}$ , Shahram Abbassi$^{1,2}$\thanks{E-mail: abbassi@ipm.ir ;}, Mohsen Shadmehri$^{3}$\\
$^{1}$Department of Physics, School of Sciences, Ferdowsi University of Mashhad, Mashhad, 91775-1436, Iran\\
$^{2}$School of Astronomy, Institute for Research in Fundamental Sciences (IPM), P.O. Box 19395-5531, Tehran, Iran\\
$^{3}$Department of Physics, Faculty of Sciences, Golestan University, Gorgan 49138-15739, Iran}
\begin{document}

\pagerange{\pageref{firstpage}--\pageref{lastpage}} \pubyear{2016}

\maketitle

\label{firstpage}
\begin{abstract}
 We obtain two-dimensional exact analytic solutions for the  structure of the hot accretion flows without wind. We assume that the only non-zero component of the stress tensor is $T_{r\varphi}$. Furthermore we assume that the value of viscosity coefficient $\alpha$ varies with $\theta$.  We find radially self-similar solutions and compare them with the numerical and the analytical solutions already studied in the literature. The no-wind solution obtained in this paper may be applied to the nuclei of some cool-core clusters.
\end{abstract}

\begin{keywords}
Accretion - accretion discs -black hole physics - hydrodynamics.
\end{keywords}
\section{Introduction}

 Hot accretion flows such as advection dominated accretion flow (ADAF) (Narayan \& Yi 1994, 1995a,b (hereafter NY94, NY95a,b); Abramowicz et al. 1995) are of great interest because they are likely operating in low luminosity active galactic nuclei, which are the majority of galaxies at least in the nearby universe, and the hard/quiescent states of black
hole X-ray binaries (see Yuan \& Narayan 2014 for the latest review). Many numerical simulations have been carried out to study the structure of hot accretion flows (e.g., Igumenshchev \& Abramowicz 1999, 2000; Stone et al. 1999, 2001; Machida et al. 2001; Hawley \& Balbus 2002; Pang et al. 2011; Yuan et al. 2012a).

 One of the most important progresses in this field is the discovery of strong wind launched from the accretion flow by numerical simulations (Yuan et al. 2012b; Narayan et al. 2012; Li et al. 2013; Sadowski et al. 2016; Bu et al. 2013; Bu et al. 2016a, 2016b). This result is confirmed by the 3 million seconds Chandra observations of the accretion flow around the super-massive black hole in the Galactic Center, combined with the modeling to the detected iron emission lines (Wang et al. 2013). Begelman (2012) and Gu (2015) analytically address the question why wind exists in hot accretion flow.

 Despite the existence of wind found by simulations of hot accretion flow, observations of the nuclei in the centers of cool-core clusters which are hot accretion flows show that outflow may be weak (Hlavecek-Larrondo \& Fabian 2011). Allen et al. (2006) studied eight massive nearby elliptical galaxies where the gas properties close to the Bondi radius can be observed or reasonably extrapolated. In all the sources, the Bondi accretion rate $\dot M_B$ can be determined by the gas properties at the Bondi radius. The jet power $P_j$ in these sources can also be determined by observations of bubbles inflated by the jet in the surrounding gas. In this case one may write $P_j=\eta \dot M_B c^2$, where $\eta$ is the jet production efficiency. It is found that $\eta\simeq 2\%$. This is a rather large efficiency. On the other hand, it is required that almost  all the gas captured at the Bondi radius by the black hole need to go into the black hole. There is almost no gas lost when the gas goes towards the black hole.
Based on the above observations, it is still necessary to find hot accretion flow solution with no wind. Many authors have find one-dimensional (e.g., NY94; Zhang \& Dai 2008; Bu et al. 2009) or two-dimensional (e.g. NY95a; Xue \& Wang 2005; Jiao \& Wu 2011; Shadmehri 2014; Gu 2015; Zeraatgari \& Abbassi 2015) analytical solutions for hot accretion flow by assuming radial self-similarity. In all the analytical solutions mentioned here, the authors assume that the viscosity coefficient $\alpha$ is a constant. However, three-dimensional numerical simulations of hot accretion flow show that $\alpha$ is not a constant, but varies with $\theta$ (Penna et al. 2013).

By adopting modified $\alpha$-prescription for viscosity, Zeratgari \& Abbassi (2015)
improved solution presented by Gu (2015). In fact they have assumed a specific latitudinal dependent form for the viscosity while keeping the $\alpha$ parameter constant. In this case the advection parameter $f$ will also vary in the latitudinal direction. However in this paper we present analytical solutions following the same methodology as described in Shadmehri (2014). In this paper, we construct a two-dimensional solution of hot accretion flow without wind. NY95a studied two-dimensional solutions of hot accretion flow without wind by assuming radial self-similarity. However, in their paper, the authors assume that the viscosity have nine components. However, numerical simulations show that the ($r,\phi$) component of viscosity dominates other components. Therefore, it is necessary to re-study the hot accretion flow structure without wind in which only the ($r, \phi$) component is presented. The solutions of hot accretion flows without wind in this paper may be applied to nuclei of cool-core clusters mentioned above.

The outline of the paper is as follows. In section 2, the governing equations in the spherical polar coordinates for a steady state flow with zero latitudinal velocity are presented. In section 3, we obtain a set of radially  self-similar solutions and discuss about their physical behavior. We then conclude with possible astrophysical implications.

\section{Basic equations}
%
%\subsection{Basic equations}
%
We solve the standard hydrodynamic equations in the spherical polar coordinates $ (r, \theta, \phi) $, where $ r $ is the radial distance, $ \theta $ and $ \phi $ are the polar and the azimuthal angles respectively.  In this study, the disc is taken to be axisymmetry (with respect to the rotational axis and the equatorial plane) and steady state $ (i.e., \partial/\partial \phi = \partial/\partial t = 0) $. Besides, the gravitational potential of the central black hole is assumed to be Newtonian, i.e. $ \psi (r) = - GM/r $. This assumption is convenient for finding radial self similar solutions. Of course this assumption means that we are studying distances which are far enough from the center. Otherwise one has to take into account the general relativistic effects. Therefore it is worth mentioning that we are using non-relativistic hydrodynamics and gravity. Moreover, we assume that the flow is not self-gravitating.  The basic equations are the continuity equation, Euler's equationnd the energy equation. The continuity equation is
\begin{equation}\label{continuity}
	\frac{1}{r^{2}}\frac{\partial}{\partial r} \left( r^{2} \rho v_{r} \right)  + \frac{1}{r \sin \theta} \frac{\partial}{\partial \theta} \left(\sin \theta \rho v_{\theta} \right) = 0,
\end{equation}
three components of the Euler's equation are
\begin{gather}\label{motion_r}
	v_{r} \frac{\partial v_{r}}{\partial r} + \frac{v_{\theta}}{r} \left( \frac{\partial v_{r}}{\partial \theta} - v_{\theta} \right) - \frac{v_{\phi}^{2}}{r} = - \frac{GM}{r^{2}} - \frac{1}{\rho} \frac{\partial p}{\partial r},\\
	v_{r} \frac{\partial v_{\theta}}{\partial r} + \frac{v_{\theta}}{r} \left( \frac{\partial v_{\theta}}{\partial \theta} +  v_{r}  \right) - \frac{v_{\phi}^{2}}{r}  \cot \theta = \frac{1}{\rho r} \frac{\partial p}{\partial \theta},\\
	v_{r} \frac{\partial v_{\phi}}{\partial r} + \frac{v_{\theta}}{r} \frac{\partial v_{\phi}}{\partial \theta} + \frac{v_{\phi}}{r} \left( v_{r} + v_{\theta} \cot \theta \right) = \frac{1}{\rho r^{3}} \frac{\partial}{\partial r} \left( r^{3} T_{r \phi} \right),
\end{gather}
and the energy equation of the gas is given by
\begin{equation}\label{energy}
	\rho \left( v_{r} \frac{\partial e}{\partial r} + \frac{v_{\theta}}{r} \frac{\partial e}{\partial \theta} \right) - \frac{p}{\rho} \left(  v_{r} \frac{\partial \rho}{\partial r} + \frac{v_{\theta}}{r} \frac{\partial \rho}{\partial \theta} \right)  = f t_{r\phi} r \frac{\partial}{\partial r} \left( \frac{v_{\phi}}{r} \right),
\end{equation}
In above equations, $  \rho $ is the gas mass density, $ v_{r} $, $ v_{\theta} $ and $ v_{\phi} $ are the three components of velocity,  $ p $ stands for the gas pressure, $ f $ refers to the advection parameter defined by NY94 and also $ e $ denotes the specific internal energy of the gas that can be expressed as,
 \begin{equation}
	e = \frac{p}{\rho \left(\gamma - 1 \right)}
\end{equation}
where $ \gamma \equiv c_{p}/c_{v} $ is the ratio of specific heats. As we mentioned before, it is assumed that only $ r\phi $-component of anomalous stress tensor is dominated. In this case it is convenient to write $ T_{r \phi}  = - \alpha p $. Following Lovelace et al. (2009), we adopt the
modified $\alpha$ description of viscosity as a function of $\theta$.

\section{Self Similar Solutions and analysis}

Following NY95a, we propose the following radially similarity solutions:
\begin{gather}
	\rho(r,\theta) = r^{-3/2} \rho(\theta) ,\\
	p(r, \theta) = r^{-5/2} GM p(\theta),\\
	v_{r}(r,\theta) = \sqrt{\frac{GM}{r}} v_{r}(\theta),\\
	v_{\theta}(r, \theta) = 0,\\
	v_{\phi}(r, \theta) =   \sqrt{\frac{GM}{r}} v_{\phi}(\theta),
\end{gather}
Upon substituting the above self-similar solutions into  Equations (\ref{continuity})-(\ref{energy}), we obtain the following algebraic equations:
\begin{gather}
	5 p(\theta) + \rho(\theta) [ v _{r}(\theta)^{2} - 2 + 2 v _{\phi} (\theta)^{2} ] = 0 ,\label{motion1}\\
	\frac{dp(\theta)}{d\theta} = \rho(\theta)  v_{\phi}(\theta)^{2}  \cot \theta ,\label{motion2}\\
	\alpha (\theta) p(\theta) + \rho(\theta) v_{r} (\theta) v_{\phi} (\theta)= 0 ,\label{motion3}\\
	(3 \gamma - 5) v_ {r} (\theta) - 3 \alpha (\theta) f (\gamma - 1) = 0 \label{energy2},
\end{gather}
From Equation (\ref{motion3}), we obtain
\begin{equation} \label{pressure_theta}
	p (\theta) = - \frac{v_{r} (\theta) v _{\phi} (\theta) \rho (\theta)}{\alpha (\theta)} ,
\end{equation}
Also from equation (\ref{energy2}), we have
\begin{equation}
 v _ {r} ( \theta) = - \frac{\alpha}{\epsilon} v_{\phi} (\theta) , \label{v_r}\\
\end{equation}
In the above equation, $ \epsilon $ is a constant defined by NY95a as follows,
\begin{equation}
	\epsilon = \frac{1}{f} \left( \frac{5/3 - \gamma}{\gamma - 1} \right),
\end{equation}
Using these equations and from equation (\ref{motion1}), the rotational velocity becomes
\begin{equation}
  v _{\phi} (\theta) = \frac{\sqrt{2} \epsilon}{\sqrt{2 \epsilon ^ 2 +5 \epsilon +  \alpha (\theta) ^ 2}} ,
\end{equation}
Then, the radial velocity is obtained  from equation (\ref{v_r}), i.e.
\begin{equation}
v_{r} (\theta) = -\frac{\sqrt{2} \alpha (\theta)}{\sqrt{2 \epsilon ^ 2 + 5 \epsilon + \alpha (\theta)^ 2}}
\end{equation}
Thus, from equation (\ref{motion2}), one can easily obtain the density of the flow,
\begin{equation}
 \rho (\theta) = \frac{\rho(\frac{\pi}{2}) \sin \theta^ {\epsilon} (2 \epsilon ^ 2 + \alpha (\theta) ^2 + 5 \epsilon)}{2 \epsilon ^2 + \alpha (\frac{\pi}{2})^ 2 + 5 \epsilon}
\end{equation}
In this case, using equation \eqref{pressure_theta} we can obtain the pressure profile
\begin{equation}\label{new2}
p(\theta)=\frac{2\epsilon}{2 \epsilon ^ 2 + \alpha (\theta) ^2 + 5 \epsilon}\rho(\theta)
\end{equation}
 Interestingly if the functional from of  $\alpha(\theta)$ is known,  then equations \eqref{motion1}-\eqref{energy2} can be integrated. In other words, the functionality of $\rho(\theta)$ can be exactly obtained for a given $\alpha(\theta)$.

Finally, the net mass accretion rate can be obtained from the volume integral of equation (\ref{continuity}) as follows,
\begin{equation}\label{mdot1}
	\dot{M} = - \int 2\pi r^{2} v_{r}(r,\theta) \rho(r,\theta) \sin\theta d \theta .
\end{equation}
Also by putting the self-similar solution into the mass accretion rate, a dimensionless form is defined as,
\begin{equation}\label{mdot2}
	\dot{m} = - \int \rho v_{r}  \sin \theta d\theta,
\end{equation}
where $ \dot{m} = \dot{M}/(2 \pi \sqrt{GM}) $. Since the radial self-similarity has been employed in this study, the radial dependency has been omitted in the above equation.  Then,  pressure at the equatorial can be determined uniquely as,

\begin{equation}
	p\left(\frac{\pi}{2}\right) = \frac{ \dot{m}}{I(\alpha, \epsilon)},
\end{equation}
where, $I(\alpha, \epsilon)$ is defined as
\begin{equation}\label{new1}
	  I(\alpha, \epsilon) = \int_{0}^{\pi} \sin \theta^{\epsilon + 1} \alpha (\theta) \sqrt{2 \epsilon^2 + 5 \epsilon + \alpha(\theta)^2} d\theta
\end{equation}
It is necessary to mention that using equation \eqref{new2}, one can find the integration constant $\rho(\frac{\pi}{2})$ as
\begin{equation}
\rho(\frac{\pi}{2})=\frac{2 \epsilon^2 + 5 \epsilon + \alpha(\frac{\pi}{2})^2}{2\epsilon}\frac{\dot{m}}{I(\alpha, \epsilon)}
\end{equation}

We can integrate equations \eqref{motion1}-\eqref{energy2} for a given $\alpha(\theta)$. However, we have found several functions for $\alpha(\theta)$ for which this integral can be solved  analytically. In the following we introduce one of them which seems more realistic and discuss its physical implications.

\subsubsection*{$\alpha(\theta)=\alpha_0\sin(\theta)$}
\label{alpha3}
Just as an illustrative example, we have chosen $\alpha=\alpha_0 \sin\theta$  where $\alpha_0$ is a constant. In fact, among our solutions, this form recovers the main features of the ADAFs presented in NY95a. On the other hand this exact solution does not have the problems of the exact solutions introduced in Shadmehri (2014). This particular form of the dimensionless stress parameter $\alpha (\theta)$ becomes zero at $\theta=0$ and reaches to a maximum at the equatorial plane of the disk. Obviously there are more functional forms for this parameter with a similar behaviour as a function of the polar angle, however, this particular choice is  the simplest case and it is also a reminiscent of the model studied by Lovelace et al. (2009) where the magneto-rotational instability (MRI) turbulence plays a key role in the dynamics of the disk. We note that Lovelace et al. (2009) has been proposed dependence of the stress parameter on the vertical coordinate of a disc in a completely different system, i.e. protoplanetary disc. But it is a good motivation to speculate dependence of the tress parameter on the spatial coordinates in ADAFs, though our knowledge about source of this possible dependence is not adequate and it needs further investigation. 

\begin{figure}
\includegraphics[width=70mm,height=80mm]{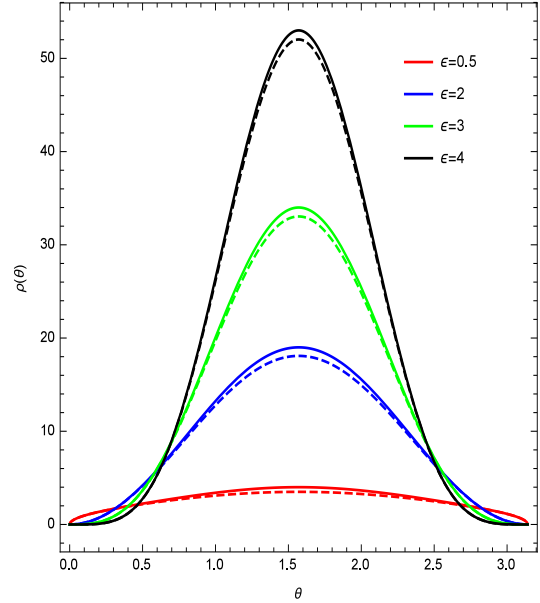}
\caption{ِDensity profile as a function of $\theta$.
Solid curves are corresponding to  $\alpha(\theta)=\alpha_0 \sin\theta$. Also dashed curves represent  Shadmehri (2014) solutions.  It is interesting that this choice for $\alpha$ leads to density profile reminiscent to that of NY94. }
\label{1}
\end{figure}

\begin{figure}
\includegraphics[width=70mm,height=80mm]{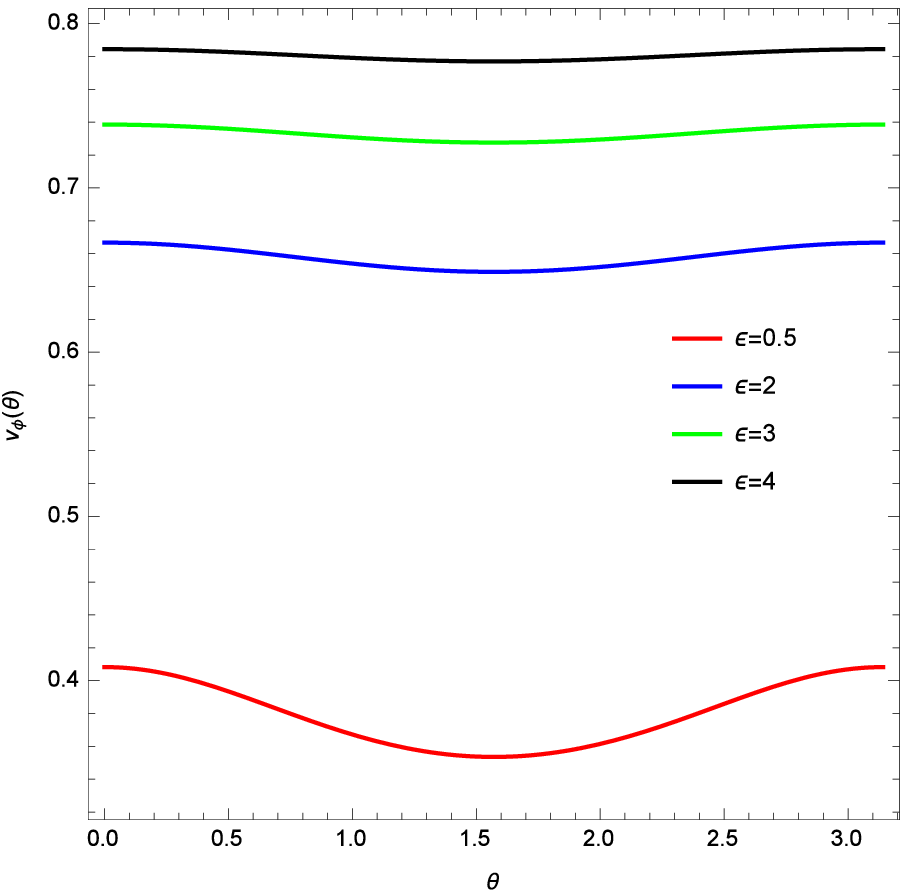}
\caption{Behavior of $v_{\phi}$ as a function of  $\theta$ for different values of $\epsilon$. }
\label{2}
\end{figure}

\begin{figure}
\includegraphics[width=70mm,height=80mm]{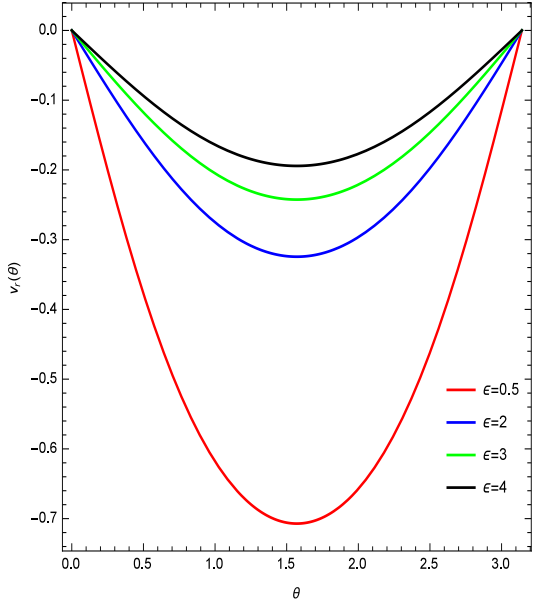}
\caption{Behavior of $v_{r}$ as a function of $\theta$ for different values of $\epsilon$.}
\label{3}
\end{figure}

We have plotted the relevant quantities, i.e. $\rho, v_{r}$ and $v_{\phi}$, in Figures \ref{1}-\ref{3}. In fact, in these figures we have chosen different values for advection parameter $f$ and a fixed value $\dot{m}=1$. Furthermore in these figures we have set the adiabatic index $\gamma$ to $1.5$. Consequently in order to check the model's behavior for different values of $\epsilon$, we have adopted different values for advection parameter. Figure \ref{1} shows that smaller values for the advection parameter leads to denser disc at $\theta=\pi/2$.

The above mentioned functionality for $\alpha$ seems interesting in the sense that the density and velocity profiles are more or less similar to those obtained in Shadmehri (2014) for a constant $\alpha$. Also there are some interesting features for this kind of viscosity. It is clear from Figure \ref{2} that the azimuthal velocity has a minimum at the plane $\theta=\pi/2$. This is consistent with the fact that the viscosity is maximum at $\theta=\pi/2$. Also it is clear that for a fixed accretion rate, the average circular velocity increases when the adevection parameter decreases. More interestingly, when the advection parameter is small, effect of  viscosity on the azimuthal velocity becomes negligible. As it is clear from Figure \ref{2}, for small advection parameter, the azimuthal velocity  $v_{\phi}$ does not change significantly with respect to the angle $\theta$. On the other hand for a larger $f$, there is an explicit minimum in $v_{\phi}$ at $\theta=\pi/2$.
Although the circular velocity for small advection parameter is almost constant with variations of $\theta$, the radial velocity $v_r$ is modified significantly even for small advection parameter $f$. Figure \ref{3} shows that the average radial velocity increases when the advection parameter increases.

\begin{figure}
\includegraphics[width=80mm,height=79mm]{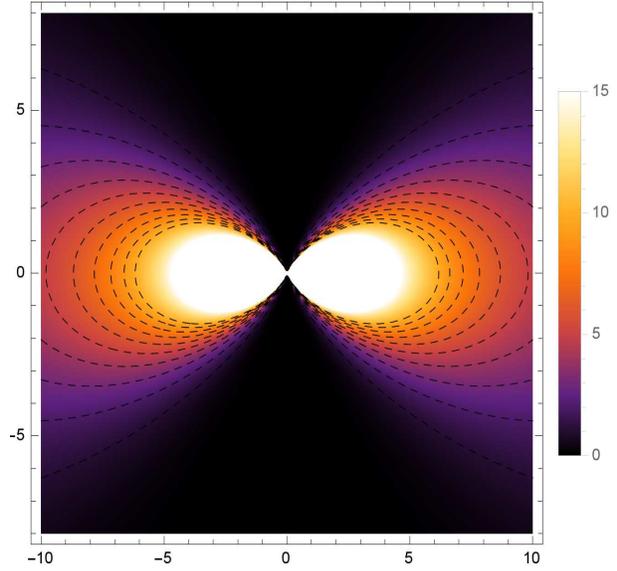}
\caption{Iso-density plot for different values of $\rho$ when $\alpha(\theta)=\alpha_0\sin(\theta)$ section \ref{alpha3}.}
\label{7}
\end{figure}

One may require more evidences from relevant hydrodynamic simulations or observations in order to justify the viability of this sinusoidal $\alpha$. As we have already shown, there is a broad variety of exact solutions. Also in order to provide a better understanding of our main exact solutions, we have plotted the iso-density profile for presented model in Figure \ref{7}.

It is instructive to find the Bernoulli parameter ($Be$) for our solution and explore the range of the advection parameter for which the Bernoulli parameter becomes positive. As it has been mentioned by NY95a, when $Be$ parameter is positive, one may expect existence of the outflows in the system. Albeit it does not mean that if $Be>0$ then necessarily there is outflow. More specifically, it can be a necessary condition for the existence of outflow and not a sufficient one. Following NY95a, let us introduce the dimensionless parameter $b$ as
\begin{equation}
b= \frac{Be}{\Omega_{\text{K}}r^2}=\frac{1}{2}v_r^2+ \frac{1}{2}(v_\phi\sin\theta)^2-1+\frac{\gamma}{\gamma-1}c_s^2,
\end{equation}
where $\Omega_{\text{K}}$ is the Keplerian angular velocity and it is defined as $\Omega_{\text{K}}= \sqrt{\frac{MG}{r^3}}$. By substituting our exact solution into this equation  for a given $\alpha(\theta)$,  we obtain
\begin{equation}\label{bernoulli}
b(\theta)=(\frac{\epsilon}{\sqrt{2\epsilon^2+5\epsilon+\alpha(\theta)^2}})^2(\sin\theta^2-2+3f)
\end{equation}

\begin{figure}
\includegraphics[width=70mm,height=80mm]{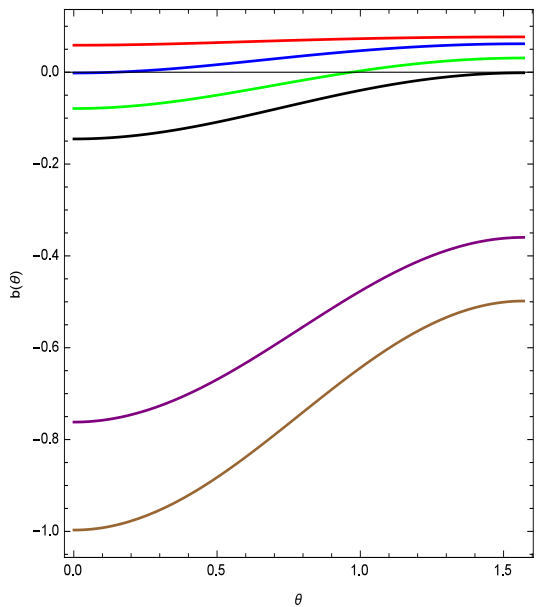}
\caption{Bernoulli parameter $b$ with respect to $\theta$ for different values of $\epsilon$. $f$, down to up in the plot, is chosen as $0.0033$, $0.033$, $0.33$, $0.44$, $0.66$, $1$. }
\label{8}
\end{figure}

The sufficient condition for positive $b>0$ is $f>2/3\simeq 0.66$. It is surprising that this constraint is independent of the form of $\alpha(\theta)$. This result is in agreement with the solutions presented in Shademhri (2014) where parameter $b$ is obtained for a constant $\alpha$. It is also important to mention that for having a positive $b$ the system should be highly advective. In Figure \ref{8} we have plotted $b$ for different values of $f$. In fact we have chosen the same values used in NY95a. It is clear that there is an obvious deviation between our analytic results and the numerical solutions of NY95a. The parameter $b$ is an increasing function with respect to $\theta$ and has its maximum at the plane $\theta=\frac{\pi}{2}$. On the other hand, in NY95a this parameter decreases as the angle $\theta$ increases. The origin for this behavior may be directly related to the behavior of the velocity components. We recall again that, as expected, the azimuthal velocity has a minimum at the plane and the magnitude of the radial component has a maximum there. In fact the maximum shear viscosity on this plane may cause a maximum radial velocity toward the center. On the other hand the slope of the radial component is much larger than the azimuthal case. Therefore, although $v_{\phi}$ decreases with $\theta$ and so does the Bernoulli parameter, the radial velocity increases more rapidly. Finally the overall behavior is that $b$ increases with $\theta$.

%\begin{figure*}
%\centering
%\includegraphics[width=170mm]{pressure.eps}
%\caption{--------- }
%\label{pressure}
%\end{figure*}
\section{discussion}

Our goal  is to provide some exact solutions for the vertical structure of the ADAFs. We assume that the only non-zero component of viscosity is $T_{r \phi}$. Moreover, the latitudinal velocity is assumed to be negligible which greatly simplifies the model.  Then, we obtained a set of self-consistent radially similarity solutions with constant accretion rate. This can also be interpreted as solutions without outflows.  Moreover, unlike analytical and semi-analytical solutions that have already presented, we assume that viscosity coefficient $\alpha$ has a prescribed  $\theta$ dependency rather than  being a constant.This assumption is consistent with numerical simulations (Penna et al. 2013). We showed that our solutions reveal most of the features of ADAF models. Also, these solutions do not have the previous problem in flattening of rotational velocity component $v_{\varphi}$ (Shadmehri 2014). Our solutions may be applied to the nuclei of some cool-core clusters where observations show that wind is weak. Our results are summarized as follows:

1. We showed that for every functionality of $\alpha(\theta)$, mass density profile of ADAFs can be calculated exactly. However the accretion rate can not be obtained for arbitrary models. For illustration, we studied a toy model which is totally integrable. For this model we found all relevant quantities and discussed their physical interpretation. Our solutions recover all general features of the ADAFs.

2. We have generalized the results of the recent paper by Shadmehri (2014). Unlike analytical solutions presented by Shadmehri (2014), our solutions include structures with latitudinal dependent azimuthal velocity. Variation of this component with respect to $\theta$ may provide more realistic situations to understand the physics of ADAFs.

3. For our toy model, we calculated the Bernoulli parameter $b$ (introduced by NY95a) and studied the conditions under which the Bernoulli parameter is positive. In this case one may expect outflows in the system. We proved that, surprisingly, independent of the form of $\alpha(\theta)$, the Bernoulli parameter would be positive if $f>0.66$.  This result is consistent with that of Shadmehri (2014) . Using the same advection parameters as in NY95a, we realized that $b$ parameter decreases with $\theta$. This result is in a gross disagreement with NY95a in which the Bernoulli parameter increases with $\theta$. Albeit it is necessary to mention that our assumptions are different from NY95a and one may naturally expect such a deviation.
\subsection*{ACKNOWLEDGMENT}
Authors would like to thank De-Fu Bu and Mahmood Roshan for valuable and constructive comments on this paper. S. Abbassi acknowledges support from the International Center for Theoretical Physics (ICTP) for a visit through the regular associateship scheme. The annoymous referee is thanked for his/her helpful comments.

\end{document}